\documentclass[aps,prl,twocolumn,showpacs]{revtex4}

\usepackage{amsmath}
\usepackage{bm}
\usepackage{graphicx}

\begin{document}
\title{Effect of Measurement on the Periodicity of the Coulomb Staircase 
of a Superconducting Box}

\author{J. M\"annik and J. E. Lukens}
\affiliation{Department of Physics and Astronomy, Stony Brook University, Stony Brook, NY11794-3800}
\date{5 November 2003}

\begin{abstract} We report on the effect of the back-action of a
Single Cooper Pair Transistor electrometer (E) on the measurement
of charge on the island of a superconducting box (B). The charge is
e-periodic in the gate bias of B when E is operated near
voltages $2\Delta/e$ or $4\Delta/e$. We show that this is due to
quasiparticle poisoning of B at a rate proportional to the number of
quasiparticle tunneling events in E per second. We are able to
eliminate this back action and recover 2e charge periodicity using a
new measurement method based on switching current modulation of E.
\end{abstract}

      
\maketitle The superconducting box, e.g. a small island of Al film
very weakly coupled to the outside circuitry by Josephson junctions,
has shown considerable promise as a qubit for quantum information
processing where the two states can be represented by superpositions
of 0 or 1 excess Cooper pairs in the box
\cite{Shnirman,Averin,Nakamura,Vion}.  Measurement of the quantum
state of this so-called charge qubit without inducing unwanted
decoherence is a significant problem as is quasiparticle poisoning,
i.e. the introduction of an unpaired electron (quasiparticle) into the
box. At temperatures of 10 mK or so, where experiments are commonly
done, the number of quasiparticles should, in principle, be
negligible.  However, such quasiparticle poisoning, due perhaps to the
measurement process itself, is commonly observed. A manifestation of
this is seen in the so called Coulomb staircase. When a charge,
$q_{g,B}$, is capacitively induced on the box, one expects Cooper pairs
to tunnel resonantly into or out of the box at $q_{g,B}=n_oe$, where
$n_o$ is an odd integer, to maintain the lowest energy charge state of
the box. This results in the Coulomb staircase of the charge in the
box $Q_{B}(q_{g,B})$ with period $2e$ in $q_{g,B}$. On the other hand,
if there are quasiparticles in the system, then maintaining the lowest
charging energy state also leads to quasiparticle tunneling. This
gives rise to splitting of the steps in the Coulomb staircase, which
shifts toward e-periodicity as the number of quasiparticles increases
\cite{Lafarge}. As a result, the lowest energy state of the box at
$q_{g,B}=n_oe$ no longer corresponds to a resonant state of the
Cooper pair tunneling. For the box qubit, this means that relaxation
does not bring the system back to its computational ground state at
its operating point. Since the ability to prepare the initial state of
the qubit is an absolutely necessary condition for quantum
computing, quasiparticle poisoning has been a serious road block for
groups working to build a charge qubit
\cite{Flees,vanderWal,Lehnert}. Solving this problem in charge qubits
is essential.
\begin{figure} [b] 
\centerline{\includegraphics[width=85mm]{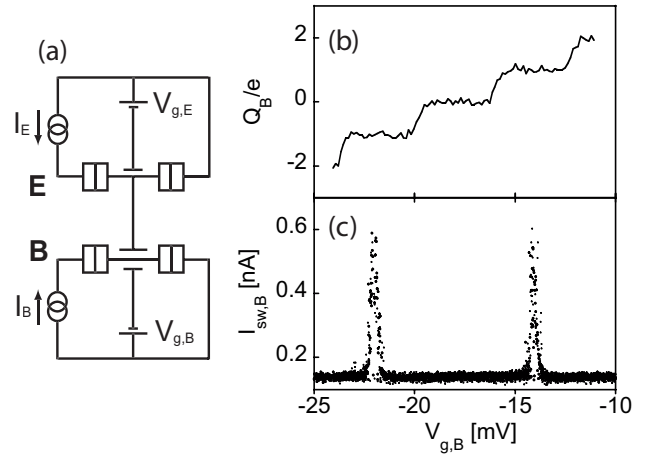}}\caption{(a)
Schematic of the measurement showing SCPT electrometer
(\textbf{E}) capacitively coupled to the second SCPT (\textbf{B}). Tunnel
junctions are represented by double box symbols. (b) Average charge
of \textbf{B} as a function of its gate voltage when electrometer
operates in VM mode. (c) Switching current of \textbf{B} as a
function of its gate voltage. There is an undetermined shift in
gate voltage between measurements in (b) and in (c) because of
drift in background charge. } \end{figure}

The purpose of this paper is to investigate the effects of measurement
(i.e.  back-action) on the measured charge in the box and to develop
approaches to minimize these effects. For this study, we use two
capacitively coupled Single Cooper Pair Transistors (SCPT)s, one of
which acts as an electrometer (\textbf{E}) and the other as a
superconducting box (\textbf{B}). The latter gives a good
representation of the box and at the same time allows us to study
quasiparticle poisoning effects without needing to operate
\textbf{E}. The SCPT electrometer can be operated in several different
modes when measuring the charge of \textbf{B}
\cite{Schoelkopf,Zorin}. Commonly, the charge measurements of
\textbf{B} are done by operating \textbf{E} in the voltage modulation
mode (VM). In this mode, \textbf{E} is biased at a sufficiently high
voltage that quasiparticles are generated, so it effectively functions
as a SET where the source drain voltage is modulated by $q_{g,E}$ with
a period of $e$. However, it is known that the switching current of a
SCPT, i.e. the current at which it switches hystereticaly from the low
voltage or phase-diffusion branch to $eV>2\Delta$, is also charge
sensitive \cite{Flees} and can be used for charge measurement. We
refer to this as the switching current mode (SW). The SW mode of
operation has been analyzed \cite{Cottet}, but until now no
measurements of the charge on the island of a box using a SW mode of
the electrometer have been reported. We present the results of the
measurements of the island charge in \textbf{B} by \textbf{E} operated
in either the VM or SW mode. The results demonstrate that
measurement-induced poisoning, which leads to an e-periodic Coulomb
staircase when using \textbf{E} in the VM mode, can be eliminated in
the SW mode.

The parameters of the sample (Fig. 1.a) are as follows: \textbf{E} has a normal
state resistance $R_{n,E} = 61.4$ k$\Omega$, charging energy $E_{c,E}
= 33$ $\mu$eV and Josephson energy $E_{J,E} = 21.2$ $\mu$eV. For
\textbf{B} these parameters are $R_{n,B} = 63.0$ k$\Omega$, $E_{c,B} =
47$ $\mu$eV and $E_{J,B} = 20.6$ $\mu$eV. $E_J$ is the average of the
Josephson energies of two junctions as determined from the values of
$R_n$ and the superconducting gap $\Delta \approx 200$ $\mu$eV by
using the Ambegaokar-Baratoff formula.  $E_c$ is determined from the
amplitude of maximum voltage modulation of the devices. The coupling
capacitance between \textbf{E} and \textbf{B} is determined to be 80
aF from their measured coupling of 4.8\%.  All devices are made using
standard two angle shadow evaporations without having normal metal
quasiparticle traps close to junctions. The sample is placed in
microwave tight copper can located on a temperature regulated stage of
a dilution refrigerator having a base temperature of 6 mK. All the
measurement leads are filtered by low temperature microwave filters
\cite{Danthesis} that are thermally anchored to the mixing chamber.
  
Figure 1.b shows the charge on the island of \textbf{B}
measured with \textbf{E} in the VM mode. During this measurement, the
source and drain leads of \textbf{B} are at a common potential with
respect to its gate. Figure 1.c presents the switching current
modulation of \textbf{B} measured with the bias current through
\textbf{E}, $I_E$, set equal to zero. In this and following
measurements, the bias current of \textbf{B}, $I_B$, is ramped at a rate
of $\approx 140$ nA/s. As can be seen, the switching current modulation of
\textbf{B} is 2e-periodic as expected at low temperature, but the
charge of \textbf{B}, measured by the electrometer, is
e-periodic. Similar dependences of $Q_B$ on $q_{g,B}$ were measured
with \textbf{E} biased in either of its voltage sensitive regions,
i.e. near the gap where $V_E \approx4\Delta/e$ or near the
Josephson-quasiparticle peak where $V_E \approx 2\Delta/e$. To
determine if the e-periodicity of $Q_B(q_{q,B})$ is due to the
back-action of \textbf{E} on \textbf{B}, the quasiparticle poisoning
rate of \textbf{B}, $\gamma_B$, was measured for a range of bias
conditions of \textbf{E}.  In addition to this, we studied how the
biases of two other SCPTs located on the same chip, but coupled more
weakly to \textbf{B}, affected $\gamma_B$.
\begin{figure}\centerline{\includegraphics[width=85mm]{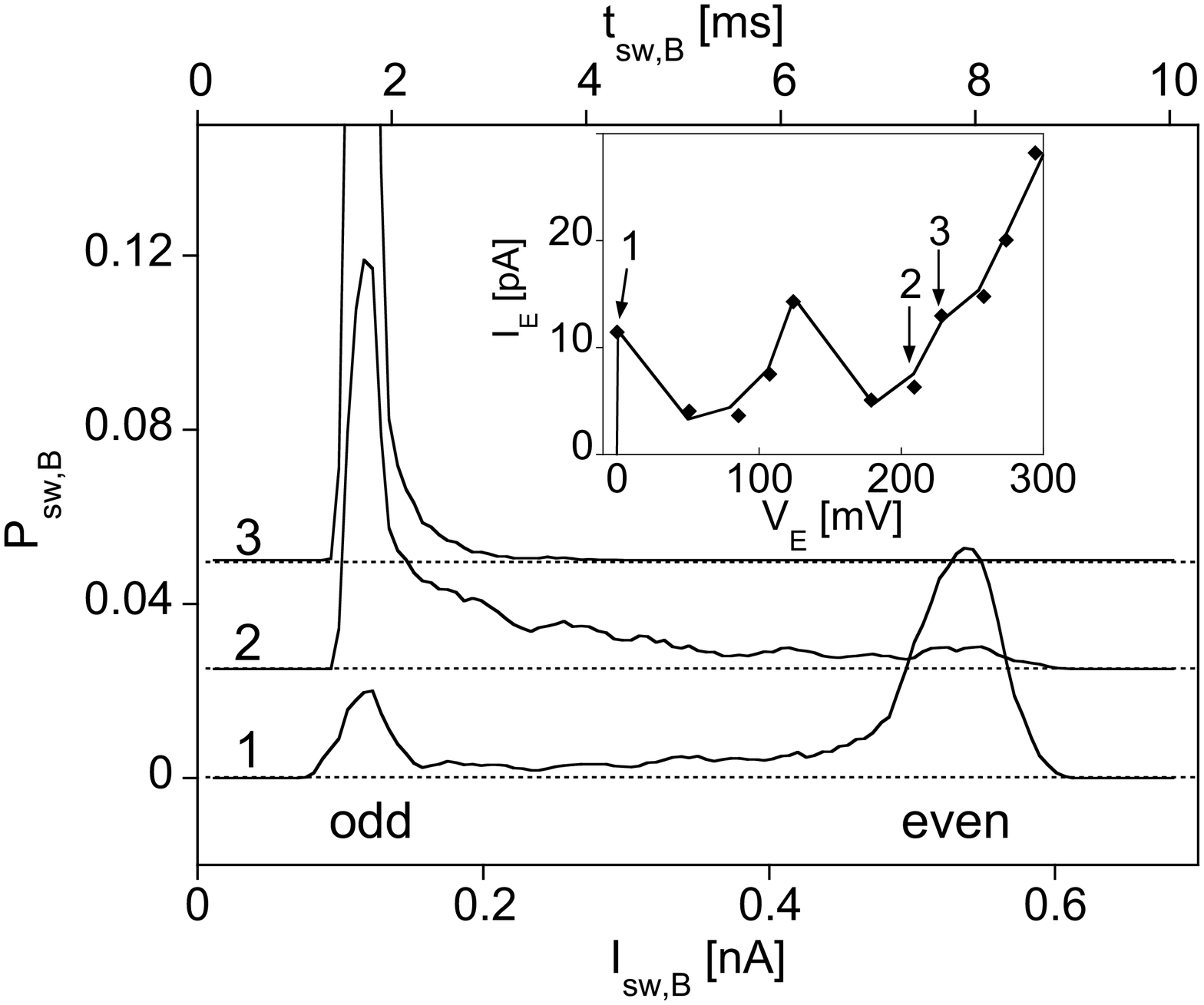}}
\caption{ Switching current histograms of \textbf{B} measured for
different bias conditions of \textbf{E}. Corresponding time delay from
the beginning of the current ramp is shown in the upper horizontal
axis. The histograms are shifted on vertical axis for clarity. The "odd"
("even") peaks in the histograms correspond to switching from the odd
(even) parity states of \textbf{B}. Inset: IV-characteristic of
\textbf{E}.  Arrows indicate bias conditions at which switching
current histograms of the main figure were measured.  Diamonds mark
positions where $\gamma_B$ has been measured for data in Fig. 3.}
\end{figure}

The details of the technique for determining $\gamma_B$ from switching
current distributions have been reported previously
\cite{Mannik}. Briefly, as $I_B$ is linearly ramped in time, the
switching current histogram of \textbf{B}, with its gate biased near
$q_{q,B}=n_oe$, exhibits two peaks if the number of quasiparticles on
the island changes during the measurement of histogram. One peak,
$I_{even}$, which is close to the maximum of the switching-current
characteristic $I_{sw,B}(q_{g,B})$, occurs when the island has an even
number of electrons (even state) and the other -much lower current-
peak, $I_{odd}$, is near the predicted switching current minimum at
$q_{g,B}=0$ when one quasiparticle occupies the island (odd state)
(Fig. 2). If $I_B>I_{odd}$ and \textbf{B} has not switched, it must be
in the even state. The entry of a quasiparticle onto the island
effectively changes $q_{g,B}=e$ to $q_{g,B}=0$ which for $I_B>I_{odd}$
will cause \textbf{B} to switch rapidly to the running state, giving
the time of the poisoning and thus the quasiparticle poisoning rate
$\gamma_B$. Previous studies have shown that $\gamma_B$ is independent
of $I_B$ in the region between the peaks, giving an exponential decay
of the even state. Several of these histograms with increasing
$\gamma_B$ as $V_E$ and $I_E$ increase are shown in Fig. 2. The
relatively low bandwidth of our filters limits these measurement to
$\gamma_B<10$ ms$^{-1}$.
      
First we determine $\gamma_B$ when all the electrodes of the electrometer
are disconnected from the measurement circuitry and grounded. In this
case we still observe a small residual rate $\gamma^0_B=0.06$
ms$^{-1}$ for $q_{g,B}\approx n_oe$.  This non-zero rate can be
caused, e.g., by the presence of impurity levels in the
superconducting gap of Al \cite{Lafarge}. For the present discussion
it is clear that this small residual rate is not related to the back-action
of \textbf{E}. $\gamma_B$ is unchanged if \textbf{E} is biased on its
supercurrent branch or when \textbf{E} is biased on its return current
branch at low voltage $V_E<200$ $\mu$V $\approx \Delta/e$ as shown in
Fig. 3. For $V_E>200$ $\mu$V, $\gamma_B$ decreases rapidly, becoming
too short to measure for slightly higher voltages. The modulation
characteristics of the current $I_E$ with gate voltage also changes at
this point from being 2e-periodic for $V_E<200$ $\mu$V to e-periodic
for $V_E>200$ $\mu$V. Similar cross-overs from e to 2e periodicity at
a voltage $\Delta/e$ have previously been seen in related systems (see
e.g. \cite{Bibow}).
\begin{figure}[t] \centerline{\includegraphics[width=85mm]{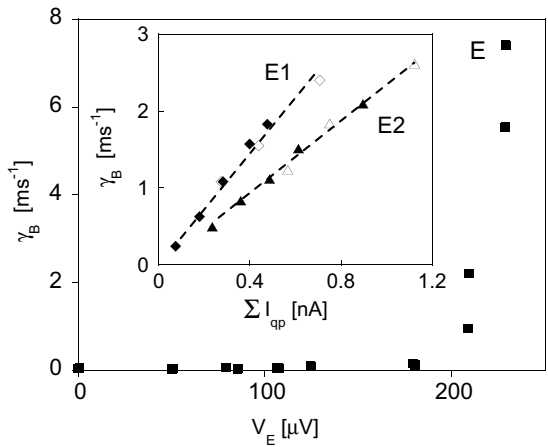}}
\caption{Quasiparticle poisoning rate $\gamma_B$ of \textbf{B} as a
function of the voltage across \textbf{E}.  Inset: $\gamma_B$ as a
function of the total quasiparticle tunneling current in
\textbf{E1} (diamonds) and of \textbf{E2} (triangles). The open and
filled symbols correspond to bias conditions $V_{E1,E2} > 4\Delta/e$
or $<4\Delta/e$, respectively. $\Sigma I_{qp} = I_{E1,2}$ or
$2I_{E1,2}$ for the filled or open symbols, respectively.}
\end{figure}

In order to study how $\gamma_B$ depends on the voltage and current of
\textbf{E} through its entire operating range, and in particular near
$V_E=2\Delta/e$ and $V_E=4\Delta/e$, we use two other SCPTs fabricated
on the same chip but much more weakly coupled to \textbf{B}. Since
these two SCPTs, which we call \textbf{E1} and \textbf{E2}, are much
more weakly coupled to \textbf{B}, they allow us to measure $\gamma_B$
for bias currents, $I_{E1,2}$, that are several orders of magnitude
higher than is possible using \textbf{E}.  The island of SCPT
\textbf{E1} is located 96 $\mu$m from the island of \textbf{B} and has
$R_n=61.4$ k$\Omega$, while the corresponding parameters for
\textbf{E2} are 143 $\mu$m and $R_n=54.2$ k$\Omega$. For these
devices, we can measure the rate $\gamma_B$ up to voltages
$4\Delta/e$. Again, we see a small initial increase of $\gamma_B$ at
$V_E\approx 200$ $\mu$V and then sharp increases at voltages
$V_E\approx2\Delta/e$ and $4\Delta/e$. These voltages correspond
approximately to the Josephson-quasiparticle tunneling and sequential
quasiparticle tunneling thresholds in a SCPT and are accompanied by
sharp increases in $I_{E1,2}$. The inset of Fig. 3 shows the rate
$\gamma_B$ as a function of the currents in \textbf{E1} and
\textbf{E2} above these two thresholds. To test the hypothesis that
$\gamma_B$ is proportional to the total number of quasiparticle
tunneling events, the currents for the electrometer voltages
$V_E>4\Delta/e$ are multiplied by two. This is done since, for a given
current at voltages $V_E>4\Delta/e$, there are twice as many
quasiparticle tunneling events through the junctions of \textbf{E} as
for $V_E\gtrsim2\Delta/e$. As one can see, this scaling collapses all
of the data from each electrometer to a common line.

From these data one can conclude that the quasiparticle current of
electrometer is the source of back-action noise leading to
quasiparticle poisoning and an e-periodic Coulomb staircase of
\textbf{B} when measured by \textbf{E}. Further, the quasiparticle
generation in \textbf{B} is proportional to the total number of
quasiparticle tunneling events per second through \textbf{E}. This
relationship could indicate that the back-action of \textbf{E} results
from the shot noise of tunneling quasiparticles. On the other hand,
this back-action could also be the result of the recombination of
quasiparticles in \textbf{E} into pairs. This quasiparticle
recombination produces phonons and to a smaller extent photons of
energy $\approx2\Delta$. These phonons/photons, which propagate from
\textbf{E} to \textbf{B} without energy relaxation could generate
quasiparticles in \textbf{B}. Determining the details of the
interaction between \textbf{E} and \textbf{B} will require further
work. However, it is interesting to note that this sort of
recombination noise would likely be suppressed by having normal metal
leads close to the junctions. This may provide an explanation of
previous results \cite{Bouchiat} in which it was possible to observe a
2e-periodic Coulomb staircase using a VM electrometer measuring a
superconducting box, which had normal metal ``quasiparticle
traps'' close to the junctions.
\begin{figure} [b] \centerline{\includegraphics[width=85mm]{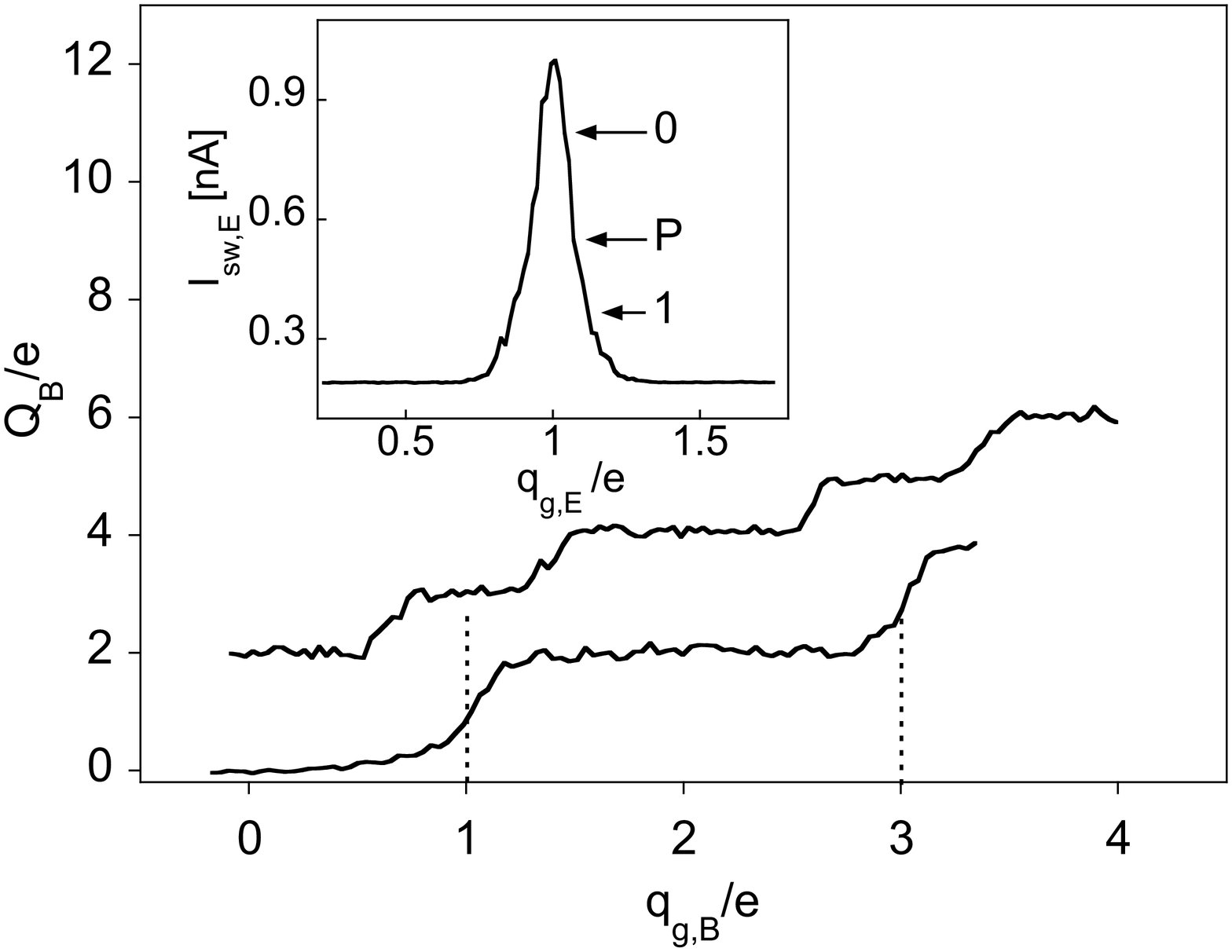}}
\caption{Average charge of \textbf{B} as a function of its gate charge
when the electrometer operates in SW mode: bottom curve, quasiparticle
flushing (see text) before each charge measurement; top curve, no
flushing. The top curve is shifted in vertical direction for
clarity. The flushing pulse has amplitude 150 $\mu$V and duration 25
ms. The dotted lines mark the positions at which the maxima of
$I_{sw,B}(q_{g,B})$ appear.  Inset shows the average switching current of
\textbf{E} as a function of its gate charge. The arrows marked by ``0'',``1''
and ``P'' correspond to switching currents of \textbf{E} when the island of
\textbf{B} has 0 or 1 excess Cooper pairs or when it is in the
poisoned state, respectively.}
\end{figure}

Quasiparticle poisoning of \textbf{B} due to the measurement of its
charge can, in principle, be eliminated by the operation of \textbf{E}
in the SW mode, where its voltage remains well below $\Delta/e$ until
the measurement is made. The SW mode of operation of the electrometer
is illustrated in the inset of Fig. 4, which shows its switching
current vs. gate charge transfer-function. In general, $q_{g,E}$
contains a component proportional to $Q_{B}$, in addition to the
externally applied bias.  So, to measure changes in $Q_B$ as a
function of $q_{g,B}$, we bias \textbf{E} near point ``0'', where the
switching current is very sensitive to variations of external charge (
inset of Fig. 4), and record several hundred switching events of
\textbf{E} for each value of $q_{g,B}$. These switching data are then
corrected for the measured nonlinearity in the transfer-function of
\textbf{E} and averaged. The top curve in Fig. 4. shows the result
obtained using this procedure.

While these data are no longer strictly e-periodic, as in Fig. 1b, the
Coulomb staircase in Fig. 4 (top line) still shows the split steps
around $q_{g,B}= n_oe$ characteristic of quasiparticle poisoning. We
can see that this is consistent with the measured residual rate,
$\gamma^0_B$, as follows. The edges of these steps should occur for an
energy difference between the even and odd states, $\Delta
E(q_{g,B})$, such that the rates for even to odd ($\gamma^0_B$) and
odd to even ($\overline{\gamma^0_B}$) transitions of the island are
equal. This ratio is given by
$\gamma^0_B/\overline{\gamma^0_B}=D_e/D_o\exp(\Delta E(q_{g,B})/kT)$
\cite{Schon}, where $D_e$ is the sum of the
quasiparticle densities on the leads and the island in the even
state. For the odd state, this sum is $D_o=D_e+V^{-1}_i$, where $V_i$
is the volume of the island. The second term in $D_o$ accounts for the
extra quasiparticle occupying the island in the odd state, which
increases the density by $V_i^{-1}\approx100$ $\mu$m$^{-3}$. $D_e$ can be
estimated from the measured residual poisoning rate $\gamma^0_B$, the
normal density of states of the Al film and the junction resistances,
giving $D_e\approx4\cdot10^{-3}$ $\mu$m$^{-3}$. Taking the electron
temperature of \textbf{B} to be 15 mK, which is reasonable since
\textbf{B} is completely passive in these measurements, gives a length
for the short step of $0.76e$ in agreement with the data shown in
Fig. 4.

The effects which result from the residual rate $\gamma^0_B$ can be
greatly reduced by flushing the quasiparticle from the island of
\textbf{B} before each measurement. As one possible approach to
prepare the even parity state, we apply a voltage pulse $V_B$ across
\textbf{B} just prior to each measurement. The amplitude of $V_B$ is
chosen such that $2E_{c,B}<V_Be<\Delta$, in order to release the
quasiparticle from the electrostatic potential of the island but yet
not to generate any new quasiparticles by the pulse
\cite{explanation}. Switching histograms of \textbf{B} show that this
procedure prepares the even state with a probability of about
85\%. Immediately after \textbf{B} is flushed, the measurement ramp of
$I_E$ begins. The result of this procedure is shown in the bottom curve
in Fig. 4. As one can see, the quasiparticle-induced splitting of the
step at $q_{q,B}=n_oe$ is no longer apparent. However, individual
histograms still show about 30\% of the switching events in \textbf{E}
near $q_{q,B}=n_oe$ are from the poisoned state of \textbf{B}. This is
consistent with the imperfect preparation of the initial state and
additional poisoning with rate $\gamma^0_B$ in the finite time between
flushing and the switching event.  Thus we see that, with the effects
of residual poisoning greatly reduced by the flushing, the measurement
of $Q_B$ by \textbf{E} in the SW mode gives results consistent with the
2e-periodic switching-current distribution of \textbf{B}.

In conclusion, our measurements clearly show that operation of a SCPT
electrometer at voltages $V_E>200$ $\mu$V causes a substantial
generation of quasiparticles in the circuit of the superconducting box
leading to an e-periodic Coulomb staircase. The rate of
quasiparticle poisoning in the box depends linearly on the total
number of quasiparticle tunneling events per second through the
junctions of the electrometer. To overcome this back-action from the
electrometer, we operate it in a mode which uses switching-current
modulation for charge detection. Using this mode of operation, we are
able to recover the 2e-periodic Coulomb staircase of the SCPT in the
box configuration, which is expected both theoretically and from the
2e-periodicity of its switching current.  

The authors thank D. V. Averin, J. R. Friedman and K. K. Likharev for
useful discussions and W. Chen and V. V. Kuznetsov for technical
assistance. Work is supported in part by AFOSR grant No. F49620010001.


\begin{thebibliography}{15}

\bibitem{Shnirman} A. Shnirman, G. Schon, and Z. Hermon,
Phys. Rev. Lett. \textbf{79}, 2371 (1997).

\bibitem{Averin} D. V. Averin, Solid State Commun. \textbf{105}, 659
(1998).


\bibitem{Nakamura} Y. Nakamura, Y. A. Pashkin, and J. S. Tsai, Nature
\textbf{398}, 786 (1999).

\bibitem{Vion} D. Vion \textit{et al.}, Science \textbf{296}, 886
(2002).

\bibitem{Lafarge} P. Lafarge \textit{et al.},
Phys. Rev. Lett. \textbf{70}, 994 (1993).

\bibitem{Flees} D. J. Flees, S. Han, and J. E. Lukens,
Phys. Rev. Lett.  \textbf{78}, 4817 (1997).

\bibitem{vanderWal} C. H. van der Wal and J. E. Mooij,
J. Supercond. \textbf{12}, 807 (1999).

\bibitem{Lehnert} K. W. Lehnert \textit{et al.},
Phys. Rev. Lett. \textbf{90}, 027002 (2003).

\bibitem{Schoelkopf} R. J. Schoelkopf \textit{et al.}, Science
\textbf{280}, 1238 (1998).

\bibitem{Zorin} A. B. Zorin, Phys. Rev. Lett. \textbf{86}, 3388 (2001).

\bibitem{Cottet} A. Cottet \textit{et al.}, in \textit{Proceedings of
Superconducting Nano-Electronic Devices}, edited by J. Pekola,
B. Ruggiero and P. Silvestrini (Kluwer Academic, New York, 2001).

\bibitem{Danthesis} D. J. Flees, {Ph.D. Thesis}, SUNY at Stony Brook,
1998.

\bibitem{Mannik} J. M\"annik \textit{et al.}, in \textit{Proceedings
of Superconducting Nano-Electronic Devices}, edited by J. Pekola,
B. Ruggiero and P. Silvestrini (Kluwer Academic, New York, 2001).

\bibitem{Bouchiat} V. Bouchiat \textit{et al.},
Phys. Scr. \textbf{T76}, 165 (1998).

\bibitem{Bibow} E. Bibow, P. Lafarge, and L. P. Levy,
Phys. Rev. Lett. \textbf{88}, 017003 (2002).

\bibitem{Schon} A similar calculation, which treats BCS quasiparticle
densities in thermal equilibrium, is in G. Schon and A. D. Zaikin,
Europhys. Lett. \textbf{26}, 695 (1994).

\bibitem{explanation} An analogous situation takes place in switching-current
measurements while a SCPT is on the return current branch so
that its voltage is $2E_c<Ve<\Delta$.

\end{thebibliography}
\end{document}